\documentstyle[aaspp4]{article}
\begin{document}
\title{On the gravitational stability of a compressed slab of gas
in the background of weakly interacting massive particles}
\author{David Tsiklauri}
\affil{Physics Department, Tbilisi State University,
3 Chavchavadze Ave., Tbilisi 380028, Georgia;
email: dtsiklau@resonan.ge}
\begin{abstract}
Linear stability of an isothermal, pressure-bounded, self-gravitating
gas slab which is gravitationally coupled with the background
weakly interacting massive particles (WIMPs) is investigated.
Analytic dispersion relations describing such a configuration are
derived. Two novel, distinct oscillatory modes are found.
Astrophysical implications of the results are discussed.
\end{abstract}

\section{Introduction}

It is widely accepted that the formation of stars occurs through
gravitational collapse of
the interstellar gas clouds. One of the possible mechanisms of
triggering the star and/or stellar cluster formation process is
a high-velocity (supersonic) cloud-cloud collisions.
In a such collision a dense gaseous slab is formed
with two plane-parallel
shock fronts propagating away from the interface (Usami et al.
1995). Then the slab grows in mass becoming unstable
against gravitational instability which causes its fragmentation.
Newly formed the gaseous clumps, in turn, collapse further
and evolve into stars and/or stellar clusters.

The structure of a self-gravitating, isothermal shock pressure-bounded
slab of gas which usually is formed in the cloud-cloud
collisions was investigated by Ledoux (1951) and later by
Elmegreen \& Elmegreen (1978) among others. Lubow \& Pringle (1993),
LP93 thereafter, have
also derived an analytic dispersion relation for the linear
perturbations and investigated, in detail, the full unstable branch
in such a slab of gas.

Possible existence of WIMP matter, one of the
form of dark matter which itself is a dominant mass component
of the universe, is strongly motivated both by
standard models of particle physics and cosmology.
It has been established that the mass density associated
with the luminous matter (stars, hydrogen clouds, x-ray gas in
clusters, etc.)
cannot account for the observed dynamics on galactic
scales and above (Trimble, 1987), therefore, revealing the existence
of large amounts of dark matter in this universe.
It has been argued that the dark matter could be
anything from novel weakly interacting elementary
particles (massive neutrinos, neutralinos) to baryonic matter
in some invisible form (brown dwarfs, primordial black holes,
cold molecular gas, etc.).

Recently, it has been shown (Tsiklauri 1998) that
classical Jeans theory of the stability of self-gravitating
interstellar gas clouds is significantly modified by
taking into account the presence of background
weakly interacting massive particles.
It has been found (Tsiklauri 1998) that the presence of
WIMP matter yields an unavoidable reduction of
the Jeans length, the Jeans mass and the Jeans time (time-scale of
the collapse via gravitational instability).

Resuming aforesaid, and in the light of recent results of Tsiklauri
(1998),
a revision of a conventional theory (e.g. LP93) of the stability of
self-gravitating, isothermal shock pressure-bounded
slab of gas against linear perturbations in the
case of presence of WIMP matter, or generally speaking any type of
{\it microscopic} non-baryonic dark matter that is coupled
with the slab only
via gravitational interaction, seems to be of a considerable
importance for full understanding of the star and/or stellar
cluster formation process.

\section{The model}

Equations that govern dynamics of two self-gravitating
fluids (a slab of gas and WIMP matter) inter-coupled only
via gravitational interaction can be written in the following
way:
$$
{{\partial \rho_i}\over{\partial t}} + \nabla (\rho_i \vec V_i)=0,
\eqno(1)
$$
$$
{{\partial \vec V_i}\over{\partial t}} +
(\vec V_i \cdot \nabla) \vec V_i = - \nabla \phi -
{{\nabla P_i}\over{\rho_i}}, \eqno(2)
$$
$$
P_i = K_i \rho_i^{\gamma_i}, \eqno(3)
$$
$$
\Delta \phi = 4 \pi G \sum_i \rho_i, \eqno(4)
$$
where notation is standard: $\vec V_i$, $P_i$, $\rho_i$ and $\phi$
denote velocity, pressure, mass density and gravitational potential
of the fluids. The subscript $i=G,W$ denotes two components,
baryonic gas and WIMP matter respectively.
Now, writing every physical quantity, for brevity commonly
denoted by $\vec f(x,z)$, in a form of
$\vec f(x,z)= \vec f^0 + \vec f^\prime(z) \exp[{i( \omega t -kx)}]$
and doing usual linearization of the Eqs.(1)--(4)
we can obtain following a set of coupled
differential equations for the perturbations
$$
{{d^2 P_G^\prime}\over{dz^2}}+
\left[{{{\omega^2}\over{C_{sG}^2}} -k^2+
{{4 \pi G \rho_G^0}\over{C_{sG}^2}}}\right]P_G^\prime=
-{{4 \pi G \rho_W^0}\over{C_{sW}^2}} P_W^\prime, \eqno(5)
$$
$$
{{d^2 P_W^\prime}\over{dz^2}}+
\left[{{{\omega^2}\over{C_{sW}^2}} -k^2+
{{4 \pi G \rho_W^0}\over{C_{sW}^2}}}\right]P_W^\prime=
-{{4 \pi G \rho_G^0}\over{C_{sG}^2}} P_G^\prime, \eqno(6)
$$
$$
{{d^2 \phi^\prime}\over{dz^2}}-k^2 \phi^\prime=
4 \pi G \left({{{P_G^\prime}\over{C_{sG}^2}}+
{{P_W^\prime}\over{C_{sW}^2}}}\right).  \eqno(7)
$$
Here, $C_{sG}$ and $C_{sW}$ denote speeds of sound for the
baryonic gas and WIMP matter respectively.
It is worthwhile to note that Eqs.(5) and (6) resemble closely
to the equations describing coupled pendulums
which arise in the non-modal study of linear perturbations
in shear flows (cf. Chagelishvili, Rogava \& Tsiklauri 1996, 1997).
The fundamental (normal) oscillatory modes of the Eqs.(5) and (6) are
given by the following expression:
$$
q_\pm^2= {{1}\over{2}}
\left[{q_G^2 +q_W^2 \pm
\sqrt{(q_G^2-q_W^2)^2+4 \alpha_G \alpha_W}}\right], \eqno(8)
$$
where
$$
q_G^2=\left[{{{\omega^2}\over{C_{sG}^2}} -k^2+
{{4 \pi G \rho_G^0}\over{C_{sG}^2}}}\right],  \eqno(9)
$$
and
$$
q_W^2=\left[{{{\omega^2}\over{C_{sW}^2}} -k^2+
{{4 \pi G \rho_W^0}\over{C_{sW}^2}}}\right],  \eqno(10)
$$
are the eigenfrequencies of the system. Also,
$\alpha_G=4 \pi G \rho_G^0/C_{sG}^2$
and $\alpha_W=4 \pi G \rho_W^0/C_{sW}^2$.

A general solution of the Eqs.(5)--(7) can be readily
found. Following LP93 we consider only symmetric modes
$dP_G^\prime/dz=dP_W^\prime/dz=0$ at $z=0$.
There are two distinct, general solutions of the Eqs.(5)--(7)
depending on the sign of $q_-^2$.

For the $q_-^2>0$ we obtain
$$
P_G^\prime = A_- \cos(q_-z)+A_+ \cos(q_+z), \eqno(11)
$$
$$
P_W^\prime = {{q_-^2-q_G^2}\over{\alpha_W}}A_- \cos(q_-z)+
{{q_+^2-q_G^2}\over{\alpha_W}}A_+ \cos(q_+z), \eqno(12)
$$
$$
\phi^\prime=C \cosh(kz)-
\left({\beta_G+{{q_-^2-q_G^2}\over{\alpha_W}} \beta_W}\right)
{{A_-}\over{k^2+q_-^2}} \cos(q_-z)-
$$
$$
\left({\beta_G+{{q_+^2-q_G^2}\over{\alpha_W}} \beta_W}\right)
{{A_+}\over{k^2+q_+^2}} \cos(q_+z), \eqno(13)
$$
here we introduced notation $\beta_G=4 \pi G / C_{sG}^2$
and $\beta_W=4 \pi G / C_{sW}^2$.

For the $q_-^2<0$ we obtain
$$
P_G^\prime = A_- \cosh(\sqrt{-q_-^2}z)+A_+ \cos(q_+z), \eqno(14)
$$
$$
P_W^\prime = {{-q_-^2-q_G^2}\over{\alpha_W}}A_-
\cosh(\sqrt{-q_-^2}z)+ {{q_+^2-q_G^2}\over{\alpha_W}}A_+
\cos(q_+z), \eqno(15)
$$
$$
\phi^\prime=C \cosh(kz)-
\left({\beta_G+{{-q_-^2-q_G^2}\over{\alpha_W}} \beta_W}\right)
{{A_-}\over{k^2-q_-^2}} \cosh(\sqrt{-q_-^2}z)-
$$
$$
\left({\beta_G+{{q_+^2-q_G^2}\over{\alpha_W}} \beta_W}\right)
{{A_+}\over{k^2+q_+^2}} \cos(q_+z), \eqno(16)
$$

Note that in the case of presence of the background WIMP matter
there are two fundamental (normal) modes $q_\pm$ (see Eq.(8)).
Whereas, in the case with no WIMP matter there is only one oscillatory
mode, namely $q_G$ (see Eq.(32) in LP93).
This is the presence of WIMP matter that causes
appearance of a novel unstable mode described by Eqs.(14)--(16).

As our next step towards derivation of the dispersion relation
for linear perturbations, following LP93, we impose relevant boundary
conditions. The first boundary condition is that the Lagrangian
pressure perturbation is zero at the slab boundary, namely that
$$
P_i^\prime + \xi_z {{dP_i^0}\over{dz}}=0, \eqno(17)
$$
at $z=a$ (with $a$ being slab half-thickness),
where $\vec \xi$ vector is the Lagrangian
displacement vector of the perturbation and $i=W,G$. In this case,
the displacement is simply related to $\vec V^\prime$ by
$\vec \xi = - i \vec V^\prime/ \omega$.
Using relation $\nabla P_i^0= - \rho_i^0 \nabla \phi^0=
-4 \pi G a \rho_i^0(\rho_G^0+ \rho_W^0)$ for the
unperturbed physical quantities, Eq. (17), at $z=a$,
reduces to the two following conditions
$$
{{P_i^\prime}\over{\rho_i^0}}= \xi_z 4 \pi G
(\rho_G^0+ \rho_W^0) a, \eqno(18)
$$

The other boundary condition is the one for
$\phi^\prime$ (LP93):
$$
{{d \phi_e^\prime}\over{dz}} - {{d \phi^\prime}\over{dz}}=
4 \pi G (\rho_G^0+ \rho_W^0) \xi_z,
$$
evaluated at $z=a$, where $\phi_e^\prime$ is the
perturbed gravitational potential in the region
$|z|>a$ and is given by
$\phi_e^\prime=\phi^\prime(a) \exp[-k(|z|-a)]$,
where it has been assumed that $k>0$.
Thus, we obtain
$$
{{d \phi^\prime}\over{dz}} \biggl|_{z=a}= -k \phi^\prime(a)
- 4 \pi G (\rho_G^0+ \rho_W^0) \xi_z. \eqno(19)
$$

\subsection{The case when $q_-^2 > 0$}

Applying boundary condistions (18) and (19) to
the general solutions in the case when $q_-^2 > 0$
[Eqs.(11)--(13)] yields
$$
{{A_+}\over{C}}=-[{\cosh(ka)+\sinh(ka)}] \times
\Biggl[{{D_+q_+}\over{k}} \sin(q_+a) -\chi {{D_-q_-}\over{k}}
\sin(q_-a)+
$$
$$
\left\{{{{1}\over{\rho_G^0ak}}-D_+}\right\} \cos(q_+a)-
\chi \left\{{{{1}\over{\rho_G^0ak}}-D_-}\right\} \cos(q_-a)
\Biggr]^{-1}. \eqno(20)
$$
Here we have introduced following notation:
$$
D_-=\left({\beta_G+{{q_-^2-q_G^2}\over{\alpha_W}} \beta_W}\right)
{{1}\over{k^2+q_-^2}},
$$
$$
D_+= \left({\beta_G+{{q_+^2-q_G^2}\over{\alpha_W}} \beta_W}\right)
{{1}\over{k^2+q_+^2}},
$$
and
$$
\chi = {{\cos(q_+a)}\over{\cos(q_-a)}}
{{[1- \delta (q_+^2 -q_G^2)/ \alpha_W]}
\over{[1- \delta (q_-^2 -q_G^2)/ \alpha_W]}}; \;\;\;
\delta \equiv {{\rho_G^0}\over{\rho_W^0}}.
$$

Another condition which relates $A_+$ and $C$
can be obtained using $z$-component of the Eq.(2) for
the perturbations in baryonic gas and
$\vec \xi = - i \vec V^\prime/ \omega$ (LP93).
Doing this yields
$$
{{A_+}\over{C}}= \rho_G^0 ak \sinh(ka)
\Biggl[
{{\omega^2 \cos(q_+a)}\over{4 \pi G(\rho_G^0+ \rho_W^0)}} -
{{\chi \omega^2 \cos(q_-a)}\over{4 \pi G(\rho_G^0+ \rho_W^0)}} -
$$
$$
 \chi q_-a(1-\rho_G^0D_-) \sin(q_-a)
+ q_+a(1-\rho_G^0D_+) \sin(q_+a)
\Biggr]^{-1}. \eqno(21)
$$
Now, combining Eqs.(20) and (21) allows us to obtain
the dispersion relation for the perturbations for
the case when $q_-^2 > 0$
$$
{{\sinh(ka)+\cosh(ka)}\over{\sinh(ka)}}=-
{{b_1^+ \cos(q_+a)+b_1^- \cos(q_-a)+b_2^+ \sin(q_+a) +b_2^-
\sin(q_-a)} \over{b_3^+ \cos(q_+a)+b_3^-
\cos(q_-a)+b_4^+ \sin(q_+a) +b_4^- \sin(q_-a)}}.
\eqno(22)
$$
where
$b_1^+=1-D_+ \rho_G^0 ak$,
$b_1^-=- \chi (1 - D_- \rho_G^0 ak)$,
$b_2^+=D_+ \rho_G^0 q_+ a$,
$b_2^-= - \chi D_- \rho_G^0 q_-a$,
$b_3^+= \omega^2 /[4 \pi G (\rho_G^0+ \rho_W^0)]$,
$b_3^- = - \chi \omega^2 / [4 \pi G (\rho_G^0+ \rho_W^0)]$,
$b_4^+=q_+a(1-\rho_G^0D_+)$,
$b_4^-=-\chi q_-a (1-\rho_G^0D_-)$.

\subsection{The case when $q_-^2 < 0$}

We now apply boundary conditions (18) and (19) to
the general solutions in the case when $q_-^2 < 0$
[Eqs.(14)--(16)]. Thus we obtain
$$
{{A_+}\over{C}}=-[{\cosh(ka)+\sinh(ka)}] \times
\Biggl[{{D_+q_+}\over{k}} \sin(q_+a) +\chi^u
{{D_-^u\sqrt{-q_-^2}}\over{k}} \sinh(\sqrt{-q_-^2}a)+
$$
$$
\left\{{{{1}\over{\rho_G^0ak}}-D_+}\right\} \cos(q_+a)-
\chi^u \left\{{{{1}\over{\rho_G^0ak}}-D_-^u}\right\}
\cosh(\sqrt{-q_-^2}a) \Biggr]^{-1}. \eqno(23)
$$
Here we have introduced following notation:
$$
D_-^u=\left({\beta_G+{{-q_-^2-q_G^2}\over{\alpha_W}}
\beta_W}\right) {{1}\over{k^2-q_-^2}},
$$
and
$$
\chi^u = {{\cos(q_+a)}\over{\cosh(\sqrt{-q_-^2}a)}}
{{[1- \delta (q_+^2 -q_G^2)/ \alpha_W]}
\over{[1- \delta (-q_-^2 -q_G^2)/ \alpha_W]}}
$$

Again using the condition which relates $A_+$ and $C$
that is obtained using $z$-component of the Eq.(2) for
the perturbations in baryonic gas and
$\vec \xi = - i \vec V^\prime/ \omega$, we get
$$
{{A_+}\over{C}}= \rho_G^0 ak \sinh(ka)
\Biggl[
{{\omega^2 \cos(q_+a)}\over{4 \pi G(\rho_G^0+ \rho_W^0)}} -
{{\chi^u \omega^2 \cosh(\sqrt{-q_-^2}a)}
\over{4 \pi G(\rho_G^0+ \rho_W^0)}} +
$$
$$
 \chi^u \sqrt{-q_-^2}a(1-\rho_G^0D_-^u) \sinh(\sqrt{-q_-^2}a)
+ q_+a(1-\rho_G^0D_+) \sin(q_+a)
\Biggr]^{-1}. \eqno(24)
$$
Finally, using Eqs.(23) and (24) we to obtain
the dispersion relation for the perturbations in
the case when $q_-^2 < 0$
$$
{{\sinh(ka)+\cosh(ka)}\over{\sinh(ka)}}=-
{{b_1^+ \cos(q_+a)+b_{1u}^- \cosh(\sqrt{-q_-^2}a)+b_2^+
\sin(q_+a) + b_{2u}^- \sinh(\sqrt{-q_-^2}a)}
\over{b_3^+ \cos(q_+a)+b_{3u}^- \cosh(\sqrt{-q_-^2}a)+
b_4^+ \sin(q_+a) +b_{4u}^- \sinh(\sqrt{-q_-^2}a)}}.
\eqno(25)
$$
where
$b_{1u}^-=- \chi^u (1 - D_-^u \rho_G^0 ak)$,
$b_{2u}^-=  \chi^u D_-^u \rho_G^0 \sqrt{-q_-^2}a$,
$b_{3u}^- = - \chi^u \omega^2 / [4 \pi G (\rho_G^0+ \rho_W^0)]$,
$b_{4u}^-= \chi^u \sqrt{-q_-^2}a (1-\rho_G^0D_-^u)$.

\section{Discussion}

In this paper we have derived two dispersions relations for
linear perturbations in a pressure-bounded, self-gravitating
gas slab which is gravitationally coupled with the
background weakly interacting massive particles.
In the original paper LP93 authors found that
the slab alone is unstable at sufficiently long wavelength and
the growth rate is of the order of $\sqrt{G\rho}$, with $\rho$
being mass density of the slab, but at high external pressures
the nature of the instability is quite different from the
classical Jeans instability. Moreover, they have developed an
analytic model which reproduces numerical results of
Elmegreen \& Elmegreen (1978). Interesting result of these authors
is that the critical wavenumber for the onset of the instability
is always of the order of the slab thickness, regardless of the
level of self-gravity and the external pressure.

Present analysis revealed that taking into account presence
of the putative WIMP dark matter significantly modifies
results of LP93 with major changes being appearance of
two distinct fundamental (normal) oscillatory modes $q_{\pm}$
and the existence of an additional unstable mode when
$q_-^2 < 0$. Obtained dispersion relations Eqs.(22) and (25)
need detailed investigation in various limiting cases.
Especially interesting would be doing analysis of the
low-frequency modes, for which LP93 found an
unexpected result that the critical wavenumber for the onset
of the instability is always of the order of the slab thickness.

\section{Acknowledgements}
I would like to thank Dr. Steven Lubow of Space Telescope Science
Institute (Baltimore) for useful comments on his previous
published work.

\end{document}